# THE LANDAU DAMPING EFFECT AND COMPLEX-VALUED NATURE OF PHYSICAL QUANTITIES

V.V. Lyahov, V.M. Neshchadim

## 1. Introduction

Complexifying physics and trying to find an opportunity of experimental verification of complex-valued nature of physical quantities [1, 2], we should not forget that two complex quantities have been already introduced into physics. Complex frequency $\omega \to \omega + i\delta$ and mass $m \to m - i\varepsilon$ were introduced. The frequency was introduced into classical physics, more exact, in plasma physics, and is associated with a so-called Landau damping in uncollisional plasma; the mass was introduced into quantum electrodynamics, in the perturbation theory. What for and how the complex frequency and mass are introduced?

In 1946 L.D. Landau's paper [3] was published as a response to the examinations initiated by A.A.Vlasov [4], [5]. A.A. Vlasov has faced the following problem: at examination of the dispersion equations for uncollisional plasma there appeared improper integrals. Velocity integrals were determined along a real axis and had a pole feature. A.A. Vlasov has substituted an improper integral by its principal value on the real axis:

$$\int_{-\infty}^{+\infty} \frac{\Phi(x)dx}{x-c} = \lim_{\varepsilon \to 0} \left\{ \int_{-\infty}^{c-\varepsilon} \frac{\Phi(x)dx}{x-c} + \int_{c+\varepsilon}^{+\infty} \frac{\Phi(x)dx}{x-c} \right\}.$$

L.D. Landau pointed out the inadmissibility of such procedure and, actually, offered to take a principal value of an integral with exit into a complex plane:

$$\lim_{v \to 0} \frac{1}{x+iv} = \frac{P}{x} - i\pi\delta(x). \tag{1}$$

This expression should be understood as its multiplication by a certain function followed by integration; imaginary $iv$ is formed by the prior introduction of the complex frequency $\omega \to \omega + i\delta$. From the mathematical point of view such procedure is correct since the complex integration could be done along any path, including the path when limits to the right and to the left of a jump of the intergrand function can be taken equal.

The formal mathematical reasons have caused us to abandon the real axis and to go to a complex plane. Thus, an essential physical result was the conclusion that uncollisional damping of electromagnetic waves in plasma is a possibility. This deduction is backed up by experimental results in the Vavilov-Cerenkov effect.

## 2. Formulation of the problem

But why is it only frequency? Why only frequency has been assigned with a small imaginary part? Is it has any distinction among other physical quantities? Let's refer to the improper integral in the Landau damping theory: s

$$\int_{-\infty}^{+\infty} \frac{\Phi(v)}{\omega - kv} dv. \qquad (2)$$

The complexification assumes that all magnitudes are to be considered as complex; the integral (2) should be presented in this case as:

$$\int_{-\infty}^{+\infty} \frac{\Phi(v)}{\omega - kv} dv = \int_{-\infty}^{+\infty} \text{Re}\left[\frac{\Phi(v)dv}{\omega - kv}\right] + i \int_{-\infty}^{+\infty} \text{Im}\left[\frac{\Phi(v)dv}{\omega - kv}\right]$$

And both addends are to be investigated. But, as it mentioned above, the most prime combinations are first be considered. Let only frequency $\omega$ and a wave vector $k$ be considered as complex.

It is seen that formally the possibility to integrate expression (2) may be provided not only through introduction of a complex frequency

$$\omega \to \omega + i\delta, \qquad (3)$$

but also the structure

$$k \to k \pm i\varepsilon. \qquad (4)$$

Let's assume that both requirements (3) and (4) are obeyed, with expression (3) be substituted by more common

$$\omega \to \omega \pm i\delta.$$

In this case:

$$\omega - kv \to (\omega \pm i\delta) - (k \pm i\varepsilon)v = \omega - kv \pm i(\delta - \varepsilon v) = x + iv.$$

If the imaginary part of the pole is $v > 0$, the formula (1) is valid; if $v < 0$, it should be used the following relation:

$$\lim_{v \to 0} \frac{1}{x - iv} = \frac{P}{x} + i\pi\delta(x). \qquad (5)$$

In selection of a positive value of the imaginary part of the complex frequency (3), we should appeal to "physical sense" which recommends to eliminate (zeroize) perturbation of the cumulative distribution function $\delta f(\mathbf{P}, \mathbf{r}, t)$ at $t = -\infty$. In the accepted time dependence $\delta f \sim exp(-i\omega t)$, such zeroing is possible in the presence of the positive imaginary part $\omega \to \omega + i\delta$. But it should be noted that first, the infinity is physically unattainable, and it is not correct to set conditions at infinity; second, in other physical theories, for example the big bang theory, it is supposed that at a start time $t_o$ the system was singular, i.e. perturbation not only did not disappear, but was infinitely large. The "physical sense" requires careful handling and is often introduced to smooth the roughness a physical theory that in general has been already created.

It should be noted that the relation (5) may be valid for frequency with positive imaginary $\omega \to \omega + i\delta$ as well, if $k \to k + i\varepsilon$ and $|\varepsilon v| > |\delta|$.

The dispersion equation for case (1) is well known, for example refer to [6]:

$$\begin{cases} 1 - \frac{\omega_{Le}^2}{\omega^2}\left(1 + 3\frac{k^2 V_{Te}^2}{\omega^2}\right) + i\sqrt{\frac{\pi}{2}} \frac{\omega \omega_{Le}^2}{k^3 V_{Te}^3} \exp\left(-\frac{\omega^2}{2k^2 V_{Te}^2}\right) = 0, \\ \left|\frac{\omega}{kV_{Ta}}\right| \gg 1, \qquad \left|\text{Re}\frac{\omega}{kV_{Ta}}\right| \gg \left|\text{Im}\frac{\omega}{kV_{Ta}}\right| \end{cases} \quad (6)$$

This equation corresponds to a high-frequency branch of longitudinal oscillations in the homogeneous isotropic plasma.

## 3. Solution of a problem

We obtained a dispersion equation for case (5) in which as opposed to (1) a particular upper point should be avoided; it takes the form:

$$\begin{cases} 1 - \frac{\omega_{Le}^2}{\omega^2}\left(1 + 3\frac{k^2 V_{Te}^2}{\omega^2}\right) - i(\sqrt{2\pi} - \sqrt{\pi/2})\frac{\omega \cdot \omega_{Le}^2}{k^3 V_{Te}^3} \exp\left(-\frac{\omega^2}{2k^2 V_{Te}^2}\right) = 0, \\ \text{when} \\ \left|\frac{\omega}{kV_{Ta}}\right| \gg 1, \qquad \left|\text{Re}\frac{\omega}{kV_{Ta}}\right| \gg \left|\text{Im}\frac{\omega}{KV_{Ta}}\right|; \\ 1 + \frac{\omega_{Le}^2}{k^2 V_{Te}^2}\left[1 - i\frac{\omega}{kV_{Te}}\left(\sqrt{2\cdot\pi}\exp\left(-\frac{\omega^2}{2k^2 V_{Te}^2}\right) - \sqrt{\pi/2}\right)\right] - \\ - \frac{\omega_{Li}^2}{\omega^2}\left[1 + 3\frac{k^2 V_{Ti}^2}{\omega^2} + i(\sqrt{2\pi} - \sqrt{\pi/2})\frac{\omega^3}{k^3 V_{Ti}^3}\exp\left(-\frac{\omega^2}{2k^2 V_{Ti}^2}\right)\right] = 0, \\ \text{when} \\ \left|\frac{\omega}{kV_{Te}}\right| \ll 1, \qquad \left|\frac{\omega}{kV_{Ti}}\right| \gg 1, \qquad \left|\text{Re}\frac{\omega}{kV_{Ti}}\right| \gg \left|\text{Im}\frac{\omega}{kV_{Ti}}\right| \end{cases} \quad (7)$$

Here

$$\omega_{Le} = \sqrt{\frac{4\pi e^2 N}{m_e}}, \quad \omega_{Li} = \sqrt{\frac{4\pi e^2 N}{m_i}}, \quad V_{Te} = \sqrt{\frac{\kappa T_e}{m_e}}, \quad V_{Ti} = \sqrt{\frac{\kappa T_i}{m_i}},$$

So, the condition (4) may result in other dispersion equation.

Let's investigate the difference between the dispersion equations (6) and (7). Radicals of the equations were determined numerically using the iterative method of Newton. It turned out that at the actual value $k = \text{Re}k$ the equation (6) is satisfied with complex frequency with the negative imaginary part $\omega = \text{Re}\omega - i\text{Im}\omega$, which determines the Landau damping. But the equation (7) (first of the two given) is satisfied with complex frequency with the positive imaginary part $\omega = \text{Re}\omega + i\text{Im}\omega$ at $k = \text{Re}k$ and at $k = \text{Re}k + i\text{Im}k$, and the positiveness of an imaginary part of frequency is testimony of an inverse process of the Landau antidamping. Table 1a shows an iterative solution of the equation (6), and Table 1b - of the equation (7). In all cases $\text{Re}k = (RL_{Debay})^{-1} = (4\pi Ne^2/\kappa T_e)^{0.5}$ where $k$ is the Boltzmann constant. Electronic plasma frequency $\omega_{Le}$ $N=10^{14}$, $T_e=T_i=10^8$ C was taken as zero approximation, which is close to the parameters of thermonuclear fusion.

## 4. Summary

If assumed that along with the frequency $\omega \to \omega + i\delta$ the wave vector $k \to k + i\varepsilon$ also has complex nature, then under certain conditions in the system (when the relation (5) is fulfilled) the Landau damping may give way to build-up of oscillation. It is possible that plasma cannot be considered merely as ionized gas (as it was supposed by A.A. Vlasov as well) described by the near- Maxwellian distribution. It may be the reason that thermonuclear fusion with magnetic confinement has not been realized till now because some modes of oscillation cannot be removed.

A similar situation is observed in the quantum electrodynamics. In order to take Feynman integrals, which are improper, the complex value is assigned to the mass $m \to m - i\varepsilon$, which is the heart of the renormalization theory. And in this case, other parameters may improve the situation along with mass.

So, the mathematical formalism of the physical theory has already caused us to enter two complex magnitudes with small imaginary parts: frequency $\omega \to \omega + i\delta$ and mass $m \to m - i\varepsilon$. Theoretical deductions prove to be true experimentally. In this situation, having put aside a paradigm about real physical quantities, it is well to leave the real axis and go to the complex plane and to explore the problems of passage to the limit in a complex neighborhood of the real axis.

Within the framework of the hypothesis offered by authors about complex-valued nature of physical quantities, the effect of the Landau damping has been explored with assumption that not only frequency can be a small imaginary component but also a wave vector. The numerical solution of the obtained dispersion equation testifies that uncollisional damping is accompanied in a certain region of space by antidumping of waves, and in particular situations antidumping may prevail over damping. It is possible that this effect may explain the experimental difficulties connected with inhibition of instabilities of plasma in the problem of controllable thermonuclear fusion.

# Appendix

Табл. 1

       а) *k*= Re*k*

ω := 564133860288.43
ω := 685019687493.094 - i·30631398795.574
ω := 783394089353.242 - i·90163039638.5647
ω := 821118276983.9     - i·153103902456.535
ω := 814444600742.901 - i·175610234350.567
ω := 812461532890.248 - i·175218623176.452
ω := 812476577012.173 - i·175217140864.446
ω := 812476577673.848 - i·175217140324.913
ω := 812476577673.848 - i·175217140324.913

       б) *k*= Re*k*
ω := 564133860288.43
ω := 685019687493.094 + i·30631398795.574
ω := 783394089353.242 + i·90163039638.5647
ω := 821118276983.9     + i·153103902456.535
ω := 814444600742.901 + i·175610234350.567
ω := 812461328890.248 + i·175218623176.452
ω := 812476577012.173 + i·175217140864.446
ω := 812476577673.848 + i·175217140324.913

       в) *k*= Re*k* + *i*Im*k*  (Jm*k*/Re*k* = $10^{**(-10)}$)

ω := 564133860288.43
ω := 685019687503.159 + i·30631398798.6953
ω := 783394089376.976 + i·90163039652.7022
ω := 821118277015.845 + i·153103902485.343
ω := 814444600774.773 + i·175610234387.84
ω := 812461532920.858 + i·175218623214.216
ω := 812476577042.796 + i·175217140902.195
ω := 812476577704.471 + i·175217140362.662